\documentclass[aps,prl,showpacs,floatfix,dblfloatfix,twocolumn,superscriptaddress,preprintnumbers,amsmath,amssymb]{revtex4-1}

\usepackage{graphicx} 
\usepackage{dcolumn}  
\usepackage{subfigure}
\usepackage{slantsc}

\usepackage{relsize}
\graphicspath{{ps}}

\newcommand{\babar}{BaBar}

\renewcommand{\arraystretch}{1.1}

\newcommand{\mevm}{\mathrm{MeV}/c^2}
\newcommand{\gev}{\,\mathrm{GeV}}

\newcommand{\gevm}{\mathrm{GeV}/c^2}

\newcommand{\ee}{e^+e^-}
\newcommand{\uu}{\mu^+\mu^-}
\newcommand{\pp}{\pi^+\pi^-}
\newcommand{\U}{\Upsilon}
\newcommand{\Uf}{\Upsilon(5S)}
\newcommand{\Uo}{\Upsilon(1S)}
\newcommand{\Un}{\Upsilon(nS)}
\newcommand{\Ut}{\Upsilon(2S)}
\newcommand{\Uth}{\Upsilon(3S)}
\newcommand{\muu}{M_{\mu^+\mu^-}}

\newcommand{\hb}{h_b(1P)}
\newcommand{\hc}{h_c(1P)}
\newcommand{\hbp}{h_b(2P)}
\newcommand{\hbn}{h_b(nP)}

\newcommand{\etabn}{\eta_b(nS)}

\newcommand{\ks}{K^0_S}

\newcommand{\fb}{\mathrm{fb}^{-1}}

\newcommand{\etal}{\em et al.}

\newcommand{\dmhf}{\Delta M_{\rm HF}}

\newcommand{\goesto}{\rightarrow}
\newcommand{\bbbar}{\mbox{$b\bar{b}$}}  

\newcommand{\jpsi}{\mbox{$ J/\psi$}}

\newcommand{\piz}{\mbox{$\pi$}^{0}}

\newcommand{\pipi}{\mbox{$\pi$}^{+}\mbox{$\pi$}^{-}}

\newcommand{\upsv}{\mbox{$\Upsilon$}{\rm (5S)}}

\newcommand{\chibnp}{\mbox{$\chi_{bJ}(nP)$}}
\newcommand{\mmpp}{\mbox{$M_{\rm miss}$}}
\newcommand{\mmppsq}{\mbox{$M^2_{\rm miss}$}}

\begin{document}

\title{\boldmath First observation of the P-wave spin-singlet bottomonium states $\hb$ and $\hbp$}

\date{August 2, 2011}
\begin{abstract}
\noindent
We report the first observation of the spin-singlet bottomonium states $\hb$ and $\hbp$ produced in the reaction $\ee\to\hbn\pp$ 
using a $121.4\,{\rm fb}^{-1}$ data sample collected at energies near the $\upsv$ resonance with the Belle 
detector at the KEKB asymmetric-energy $\ee$ collider.
We determine $M[\hb]=(9898.3\pm1.1^{+1.0}_{-1.1})\,\mevm$ and
$M[\hbp]=(10259.8\pm0.6^{+1.4}_{-1.0})\,\mevm$, which correspond to $P$-wave hyperfine splittings
$\dmhf=(+1.6\pm1.5)\,\mevm$ and $(+0.5^{+1.6}_{-1.2})\,\mevm$,
respectively.  The $\hb$ and $\hbp$ are observed with significances of $5.5\,\sigma$ and $11.2\,\sigma$, respectively. 
We also report measurements of the cross sections for $\ee\to\hbn\pp$ 
relative to that for $\ee\to\Ut\pp$. 
\end{abstract}

\pacs{14.40.Pq, 13.25.Gv, 12.39.Pn}

\affiliation{Budker Institute of Nuclear Physics SB RAS and Novosibirsk State University, Novosibirsk 630090}
\affiliation{Faculty of Mathematics and Physics, Charles University, Prague}
\affiliation{University of Cincinnati, Cincinnati, Ohio 45221}
\affiliation{Justus-Liebig-Universit\"at Gie\ss{}en, Gie\ss{}en}
\affiliation{Gifu University, Gifu}
\affiliation{Hanyang University, Seoul}
\affiliation{University of Hawaii, Honolulu, Hawaii 96822}
\affiliation{High Energy Accelerator Research Organization (KEK), Tsukuba}
\affiliation{Hiroshima Institute of Technology, Hiroshima}
\affiliation{Indian Institute of Technology Guwahati, Guwahati}
\affiliation{Indian Institute of Technology Madras, Madras}
\affiliation{Institute of High Energy Physics, Chinese Academy of Sciences, Beijing}
\affiliation{Institute of High Energy Physics, Vienna}
\affiliation{Institute of High Energy Physics, Protvino}
\affiliation{INFN - Sezione di Torino, Torino}
\affiliation{Institute for Theoretical and Experimental Physics, Moscow}
\affiliation{J. Stefan Institute, Ljubljana}
\affiliation{Kanagawa University, Yokohama}
\affiliation{Institut f\"ur Experimentelle Kernphysik, Karlsruher Institut f\"ur Technologie, Karlsruhe}
\affiliation{Korea Institute of Science and Technology Information, Daejeon}
\affiliation{Korea University, Seoul}
\affiliation{Kyungpook National University, Taegu}
\affiliation{\'Ecole Polytechnique F\'ed\'erale de Lausanne (EPFL), Lausanne}
\affiliation{Faculty of Mathematics and Physics, University of Ljubljana, Ljubljana}
\affiliation{Luther College, Decorah, Iowa 52101}
\affiliation{University of Maribor, Maribor}
\affiliation{Max-Planck-Institut f\"ur Physik, M\"unchen}
\affiliation{University of Melbourne, School of Physics, Victoria 3010}
\affiliation{Nagoya University, Nagoya}
\affiliation{Nara Women's University, Nara}
\affiliation{National Central University, Chung-li}
\affiliation{National United University, Miao Li}
\affiliation{Department of Physics, National Taiwan University, Taipei}
\affiliation{H. Niewodniczanski Institute of Nuclear Physics, Krakow}
\affiliation{Nippon Dental University, Niigata}
\affiliation{Niigata University, Niigata}
\affiliation{University of Nova Gorica, Nova Gorica}
\affiliation{Osaka City University, Osaka}
\affiliation{Pacific Northwest National Laboratory, Richland, Washington 99352}
\affiliation{Panjab University, Chandigarh}
\affiliation{Research Center for Nuclear Physics, Osaka}
\affiliation{University of Science and Technology of China, Hefei}
\affiliation{Seoul National University, Seoul}
\affiliation{Sungkyunkwan University, Suwon}
\affiliation{School of Physics, University of Sydney, NSW 2006}
\affiliation{Tata Institute of Fundamental Research, Mumbai}
\affiliation{Excellence Cluster Universe, Technische Universit\"at M\"unchen, Garching}
\affiliation{Tohoku Gakuin University, Tagajo}
\affiliation{Tohoku University, Sendai}
\affiliation{Department of Physics, University of Tokyo, Tokyo}
\affiliation{Tokyo Institute of Technology, Tokyo}
\affiliation{Tokyo Metropolitan University, Tokyo}
\affiliation{Tokyo University of Agriculture and Technology, Tokyo}
\affiliation{CNP, Virginia Polytechnic Institute and State University, Blacksburg, Virginia 24061}
\affiliation{Yonsei University, Seoul}
  \author{I.~Adachi}\affiliation{High Energy Accelerator Research Organization (KEK), Tsukuba} 
  \author{H.~Aihara}\affiliation{Department of Physics, University of Tokyo, Tokyo} 
  \author{K.~Arinstein}\affiliation{Budker Institute of Nuclear Physics SB RAS and Novosibirsk State University, Novosibirsk 630090} 
  \author{D.~M.~Asner}\affiliation{Pacific Northwest National Laboratory, Richland, Washington 99352} 
  \author{T.~Aushev}\affiliation{Institute for Theoretical and Experimental Physics, Moscow} 
  \author{T.~Aziz}\affiliation{Tata Institute of Fundamental Research, Mumbai} 
  \author{A.~M.~Bakich}\affiliation{School of Physics, University of Sydney, NSW 2006} 
  \author{E.~Barberio}\affiliation{University of Melbourne, School of Physics, Victoria 3010} 
  \author{V.~Bhardwaj}\affiliation{Panjab University, Chandigarh} 
  \author{B.~Bhuyan}\affiliation{Indian Institute of Technology Guwahati, Guwahati} 
% ...
\collaboration{The Belle Collaboration}

\maketitle

{\renewcommand{\thefootnote}{\fnsymbol{footnote}}}
\setcounter{footnote}{0}

Bottomonium is the bound system of $\bbbar$ quarks and is considered 
an excellent laboratory to study Quantum Chromodynamics (QCD) at low energies. 
The system is approximately non-relativistic due to the large $b$ quark mass, 
and therefore the quark-antiquark QCD potential can be investigated via $\bbbar$ spectroscopy~\cite{QWG}.

The spin-singlet states $\hbn$ and $\etabn$ alone provide information 
concerning the spin-spin (or hyperfine) interaction in bottomonium. 
Measurements of the $\hbn$ masses provide 
unique access to the $P$-wave hyperfine splitting, $\dmhf\equiv\langle M(n^3P_J)\rangle -M(n^1P_1)$, 
the difference between the spin-weighted average mass of the $P$-wave triplet states ($\chi_{bJ}(nP)$ or $n^3P_J$)
and that of the corresponding $\hbn$, or $n^1P_1$.  
These splittings are predicted to be close to zero~\cite{godros}, and 
recent measurements of the $\hc$ mass correspond to 
a $P$-wave hyperfine splitting that validates this expectation for the 1P level in charmonium: 
$\dmhf=(0.00\pm0.15)\,\mevm$~\cite{hcmass}.  

Recently, the CLEO Collaboration observed the process $\ee\to\hc\pipi$ at a rate comparable to
that for $\ee\to\jpsi\pipi$ in data taken above open charm threshold~\cite{cleo_hcpipi}.  
Such a large rate was unexpected because the production of $\hc$ requires a $c$-quark spin-flip, 
while production of $\jpsi$ does not.  Similarly, the Belle Collaboration observed anomalously high rates for
$\ee\goesto\U(nS)\pp$ ($n=1,2,3$) at energies near the $\Uf$ mass~\cite{5s_rate}. Together, these
observations motivate a search for $\ee\goesto\pipi\hbn$ above open-bottom threshold at the $\Uf$ resonance.

In this Letter, we report the first observation of the $\hb$ and
$\hbp$ produced via $\ee\to \hbn\pp$ in the $\Uf$ region. We use 
a $121.4\,\fb$ data sample collected
near the peak of the $\Uf$ resonance ($\sqrt{s}\sim 10.865\gev$) with the Belle detector~\cite{BELLE_DETECTOR} at the
KEKB asymmetric-energy $\ee$ collider~\cite{KEKB}.  

We observe the $\hbn$ states in the $\pp$ missing mass spectrum of hadronic events.  
The $\pp$ missing mass is defined as
\(\mmppsq\equiv (P_{\Uf} - P_{\pp})^2,\)
where $P_{\Uf}$ is the 4-momentum of the $\Uf$ determined from the 
beam momenta and $P_{\pp}$ is the 4-momentum of the $\pp$
system.  The $\pp$ transitions between $\U(nS)$ states provide
high-statistics reference signals.

\begin{figure}[tb]
\includegraphics[width=6cm]{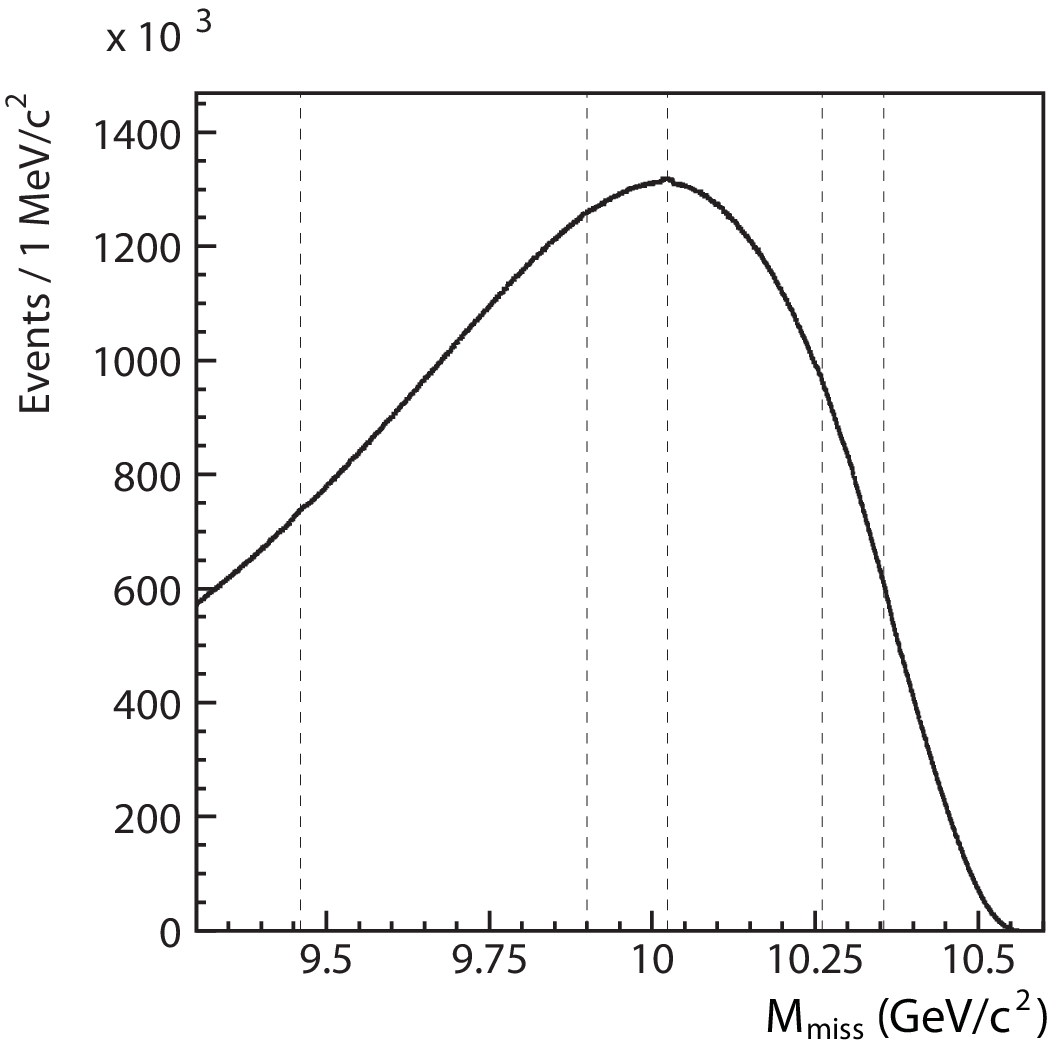}
\caption{ The $\mmpp$ distribution for the selected $\pp$
  pairs. Vertical lines indicate the locations of the $\Uo$, $\hb$,
  $\Ut$, $\hbp$ and $\Uth$ signals. }
\label{mmpp}
\end{figure}

Our hadronic event selection requires a 
reconstructed primary vertex consistent with the run-averaged interaction
point (IP), at least three high-quality charged tracks, 
a total visible energy 
greater than $0.2\,\sqrt{s}$, a total neutral energy of $(0.1-0.8)\,\sqrt{s}$, more than one 
large-angle cluster in the electromagnetic calorimeter and 
that the total center-of-mass momentum have 
longitudinal component smaller than $0.5\,\sqrt{s}$~\cite{hadronbj}.
The $\pp$ candidates are
pairs of well reconstructed, 
oppositely charged tracks that are 
identified as pions and do not satisfy electron-identification criteria.   
Continuum $\ee\to q\bar{q}$ ($q=u,\;d,\;s,\;c$) 
background is suppressed by requiring 
the ratio of the second to zeroth Fox-Wolfram moments to satisfy $R_2<0.3$~\cite{Fox-Wolfram}. 
The resulting $\mmpp$ spectrum, which is dominated by combinatoric $\pp$ pairs, 
is shown in Fig.~\ref{mmpp}.

\begin{figure*}[!]
\centering
\includegraphics[width=0.2725\linewidth]{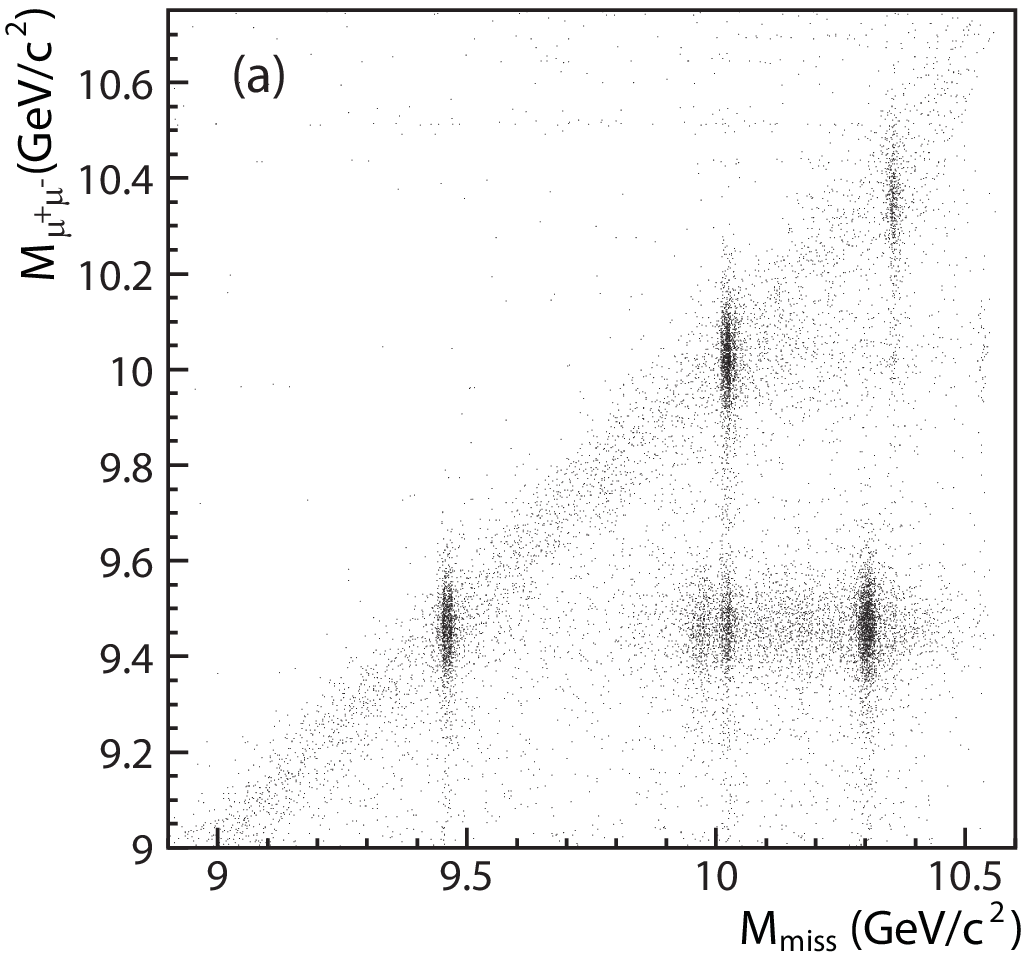}
\includegraphics[width=0.275\linewidth]{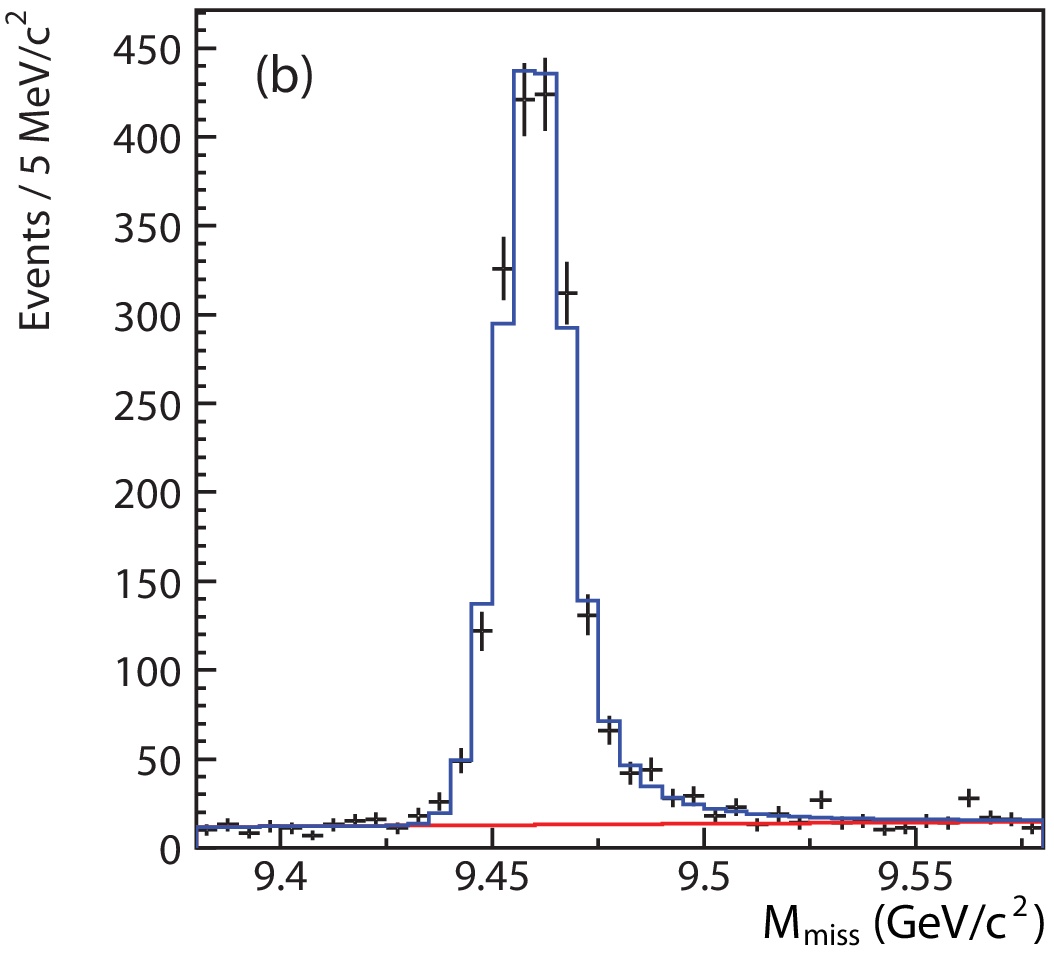}
\includegraphics[width=0.275\linewidth]{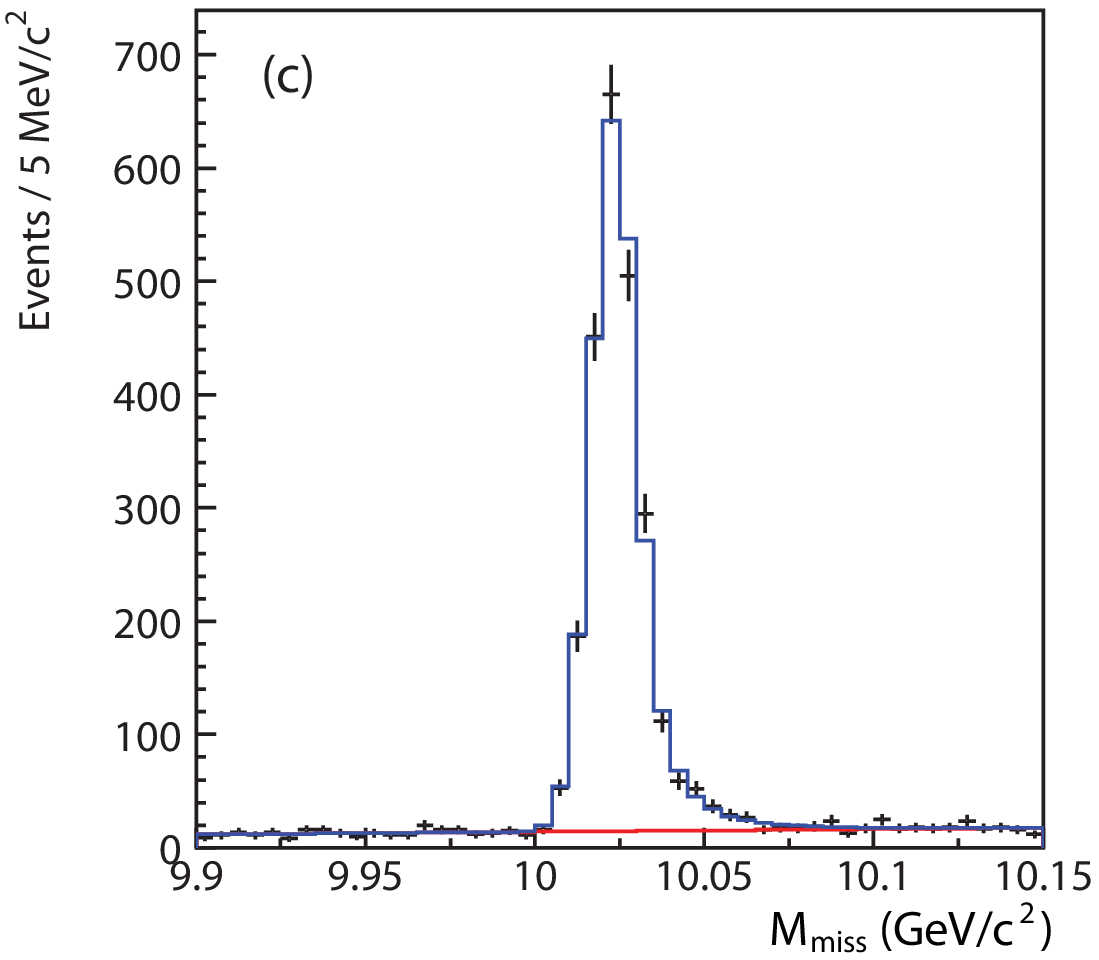}

\includegraphics[width=0.275\linewidth]{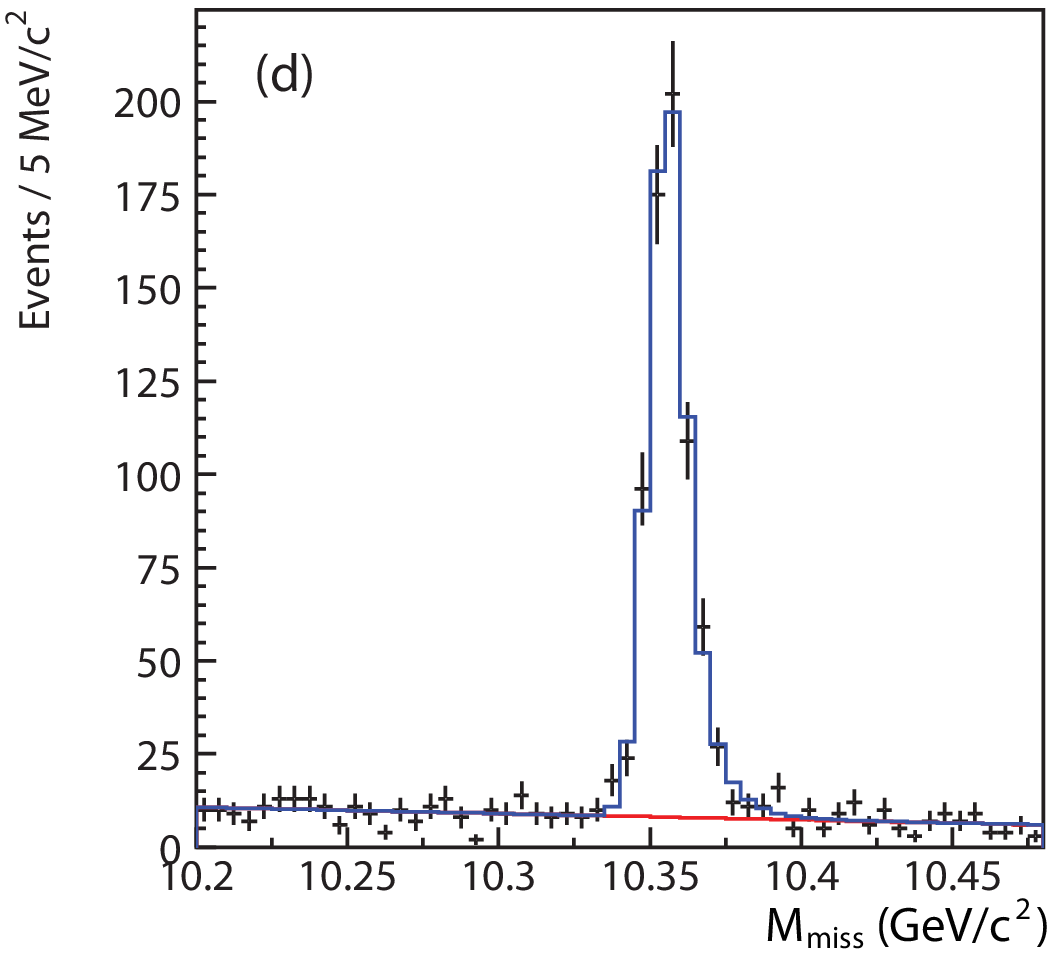}
\includegraphics[width=0.275\linewidth]{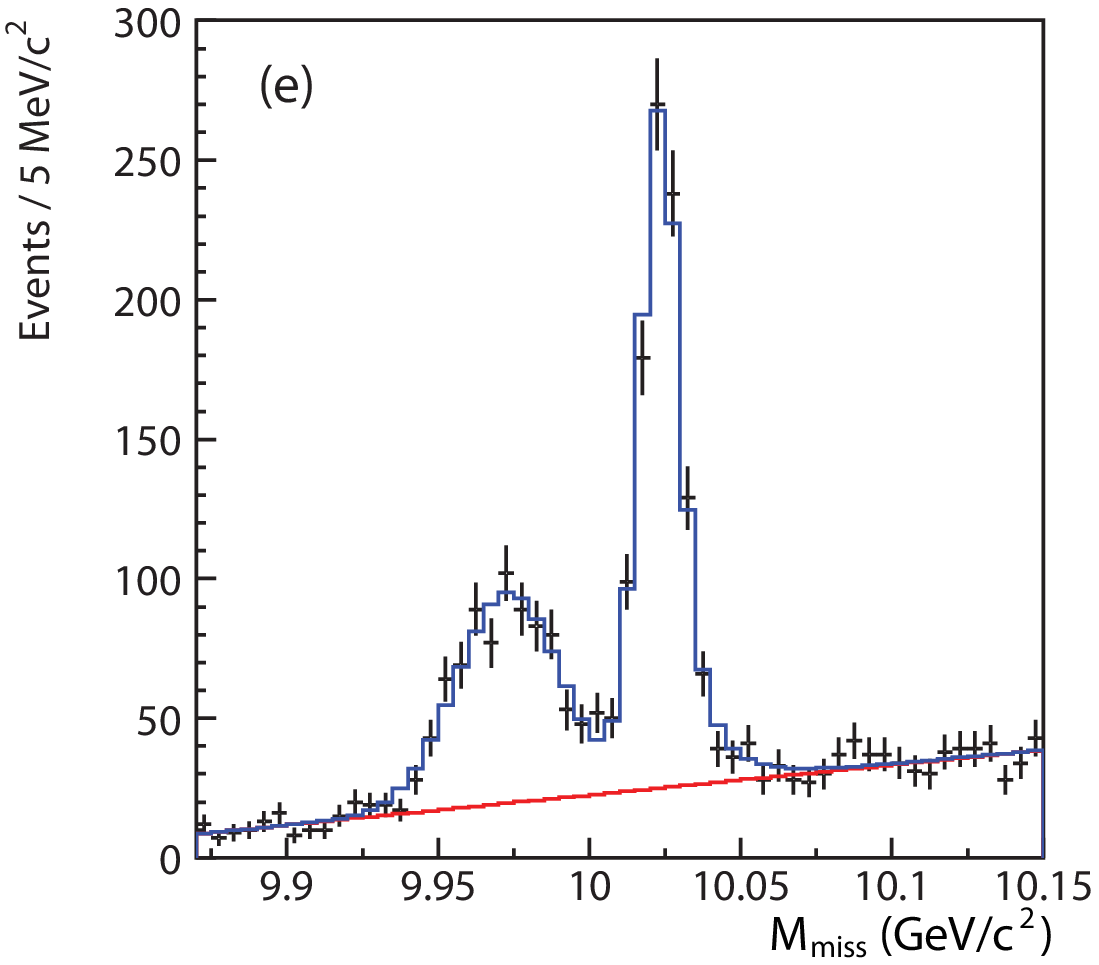}
\includegraphics[width=0.275\linewidth]{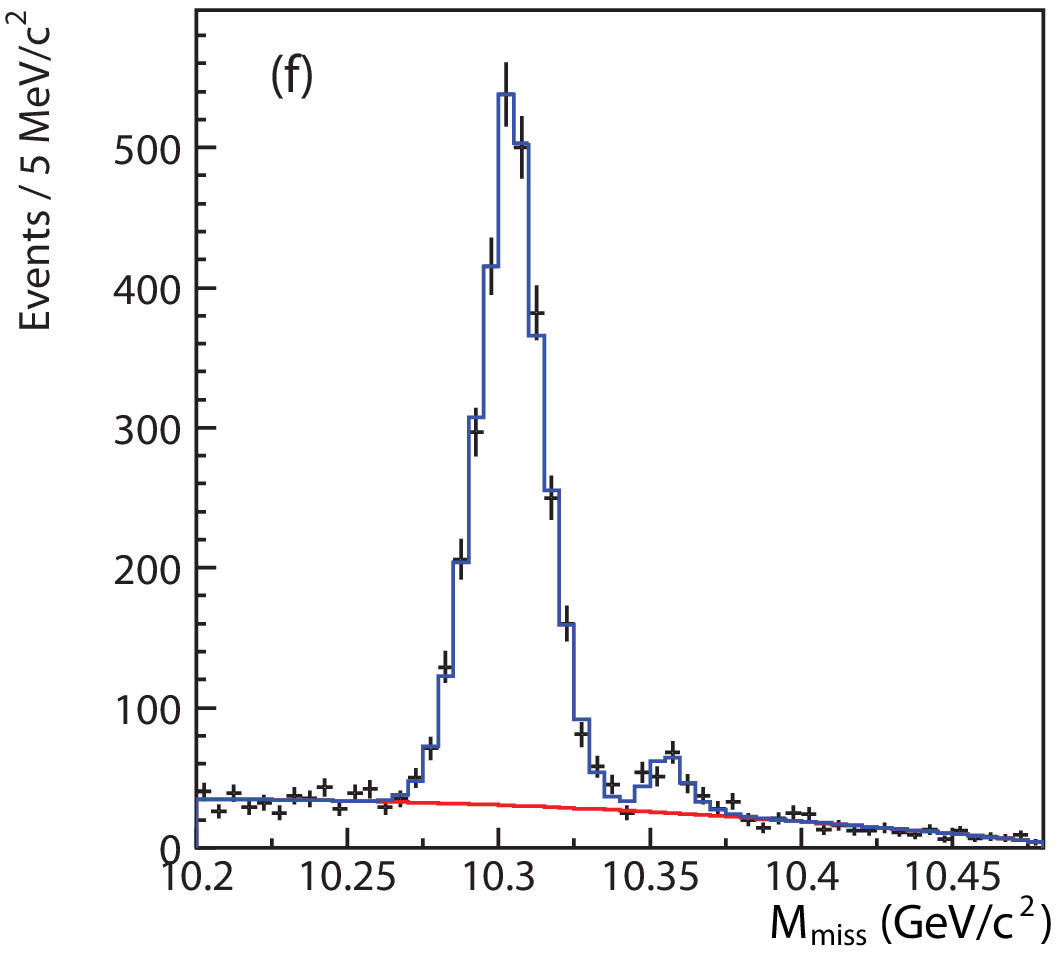}
\caption{ (a) Distribution of $\muu$ vs. $\mmpp$, and the projection on
  $\mmpp$ from (b)-(d), the diagonal band $|\mmpp-\muu|<150\,\mevm$ near the $\Uo$, $\Ut$ and $\Uth$; 
  and (e),(f), the  horizontal band $|\muu-M[\Uo]|<150\,\mevm$   near
  the $\Ut$ and $\Uth$.}
\label{fig_exc}
\end{figure*}

Prior to fitting the inclusive $\mmpp$ spectrum we study reference channels and
peaking backgrounds arising from $\pp$ transitions between $\Un$ states. A high
purity sample of such transitions is obtained by reconstructing $\uu$ pairs 
in the event in addition to the $\pp$ pair.  For these studies the hadronic event selection
criteria are not applied, while for the $\uu$ pair we use the same selection as was employed in Ref.~\cite{5s_rate}. 
MC studies indicate that the shape of the peaks in $\mmpp$ is independent of whether the $\pp$ are reconstructed 
in the hadronic environment or in this much cleaner environment.  
In addition, to suppress radiative Bhabha events in which the photon converts, producing a fake $\pipi$, 
we require that the opening angle between the candidate pions in the laboratory frame satisfies $\cos\theta_{\pp}<0.95$.  
In Fig.~\ref{fig_exc}~(a) we present 
the two-dimensional distribution of $\uu$ mass $\muu$ vs. $\mmpp$ for events satisfying these criteria.
%

Clear peaks are visible along a diagonal band, where $\muu$ is roughly equal to $\mmpp$, 
and correspond to fully reconstructed 
$\Uf\to\Upsilon(nS)\pp\to\uu\pp$ events.  Also along the diagonal is a diffuse background of events 
that arise due to the process $\ee\to\uu\gamma(\to\ee)$, where the conversion pair is reconstructed as $\pp$, 
or from non-resonant $\ee\to\uu\pp$ events.
Events from the band satisfying $|\mmpp-\muu|<150\,\mevm$
are projected onto the $\mmpp$ axis and
fitted to the sum of a linear background and a
Gaussian joined to a power-law tail on the high mass side. The high-side tail 
is due to Initial State Radiation (ISR) photons.  
This latter function is analogous to the well-known Crystal
Ball function~\cite{skwarthesis} but has the tail on the higher rather
than lower side. We thus refer to it as a 'reversed Crystal
Ball' (rCB) function.  
The fitted $\mmpp$ spectra from this band are shown in Figs.~\ref{fig_exc}~(b)-(d), 
and the resulting yields, masses and width of the rCB function for the $\Un$ states are 
displayed in Table~\ref{tab_slanted}.  The masses obtained are  
consistent with the world average values~\cite{PDG}.   

\begin{table}[tb]
\caption{ The yield, mass and width for signals reconstructed using $\mmpp$ 
from the  
exclusive $\uu\pp$ 
selection.  
  Each mass is consistent with the world average~\cite{PDG}. }
\label{tab_slanted}
\renewcommand{\arraystretch}{1.2}
\begin{ruledtabular}
\begin{tabular}{c|lll}
& Yield & Mass, $\mevm$ & $\sigma$, $\mevm$ \\
\hline
$\U(1S)$ & $1894\pm61$ & $9459.96\pm0.23$ & $7.68\pm0.21$ \\
$\U(2S)$ & $2322\pm60$ & $10023.34\pm0.22$ & $6.60\pm0.20$ \\
$\U(3S)$ & $661^{+39}_{-30}$ & $10355.66^{+0.56}_{-0.39}$ & $5.98^{+0.62}_{-0.37}$
\end{tabular}
\end{ruledtabular}
\end{table}
The structures in the horizontal band in Fig.~\ref{fig_exc}~(a), where $\muu$ is roughly equal to $M[\Uo]$, 
arise from events
in which a daughter $\Uo$ in the event decays to $\uu$.  
In Figs.~\ref{fig_exc}~(e)-(f) we present 
$\mmpp$ projections from this band, subject to the requirement $|\muu-M[\U(1S)]|<150\,\mevm$. 
The peaks at the $\Uth$ and $\Ut$ masses 
arise from events having $\pp$ transitions to $\Uth$ or $\Ut$, followed by inclusive 
production of $\Uo$, and are fitted to rCB functions.
Peaks at $9.97\,\gevm$ and $10.30\,\gevm$ arise from events in which 
a $\Uth$ or $\Ut$ is produced inclusively in $\Uf$ decays or via ISR, and then decays to
$\Uo\pp$, and are fitted to single and double Gaussians, respectively. 

The threshold for inclusive $\ks$ production results in a sharp rise in the $\mmpp$ spectrum,
due to $\ks\to\pp$, very close to the 
mass of $\Uth$.  Rather than veto $\pp$ combinations with invariant masses near $M(\ks)$, 
which significantly distorts the $\mmpp$ spectrum in the vicinity, 
we obtain the $\ks$ contamination by fitting the $\pp$ invariant mass 
corresponding to bins of $\mmpp$.  

\begin{figure*}[t]
\hspace*{-.5cm}
\includegraphics[width=0.9\linewidth]{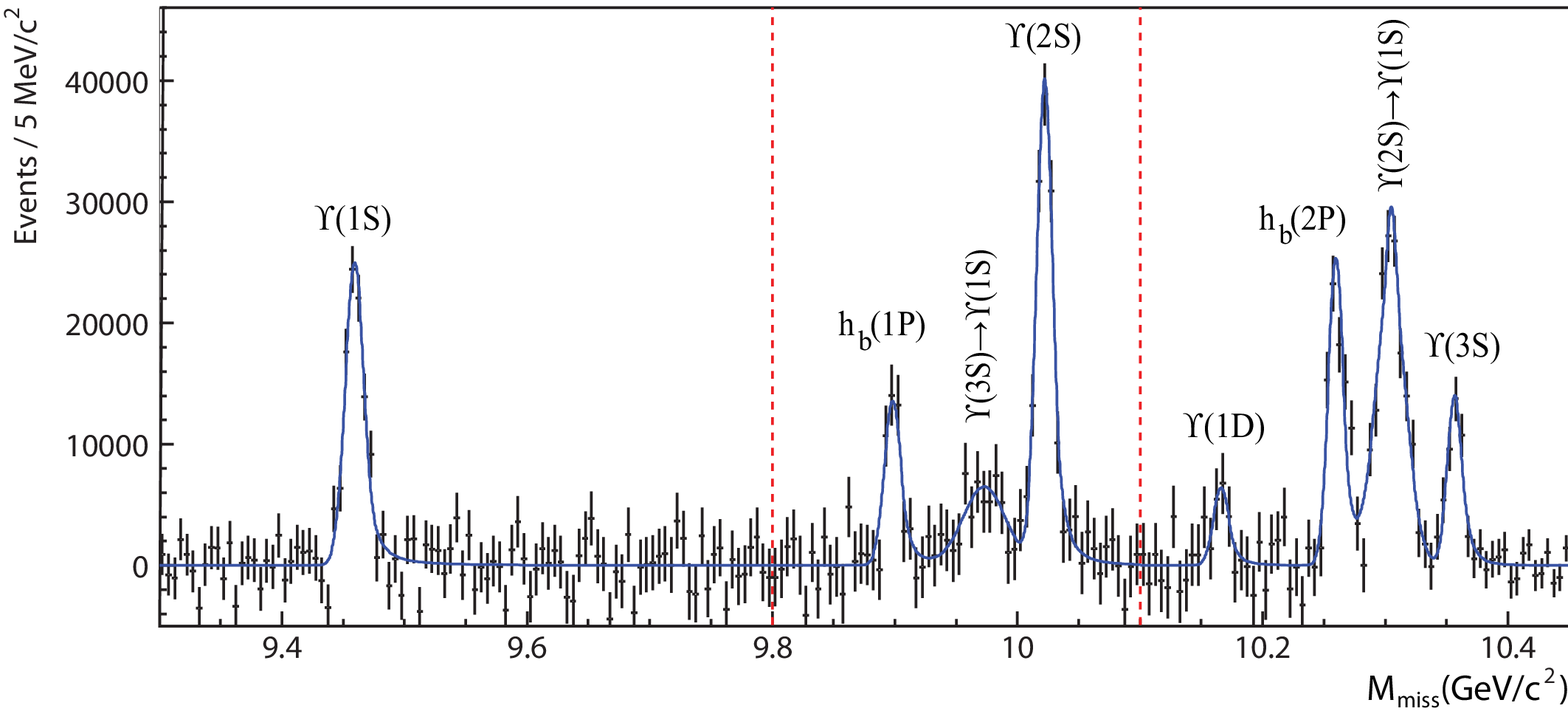}
\caption{ The inclusive $\mmpp$ spectrum with the combinatoric background and
  $\ks$ contribution subtracted (points with errors) and signal
  component of the fit function overlaid (smooth curve).  The vertical
  lines indicate boundaries of the fit regions. 
   }
\label{mmpp_all}
\end{figure*}

The $\mmpp$ spectrum is divided into three adjacent regions
with boundaries at $\mmpp=9.3$, $9.8$, $10.1$ and
$10.45\,\gevm$ and fitted separately in each region.  
In the first two regions, we use a 6th-order Chebyshev polynomial, while in the third we use a 7th-order one.
In the third region, prior to fitting, we subtract the contribution due to $\ks\to\pp$ bin-by-bin.
The signal component of the fit includes all signals seen in the $\uu\pp$ data as well as those arising from 
$\pp$ transitions to $\hbn$ and $\U(1D)$.  We fit these additional signals 
using the tail parameters of the $\Ut$ and 
fixed widths found by linear interpolation in mass from the widths of the exclusively-reconstructed $\U(nS)$ peaks.
The peak positions of all signals are floated, except that for 
$\Uth\to\Uo\pp$, which is poorly constrained by the fit.  The confidence levels of the fits in the three regions are 
$3.0\%$, $0.5\%$ and $0.4\%$, respectively. 
\begin{table}[tb!h]
\caption{The yield, mass and statistical significance from the fits to
  the $\mmpp$ distributions. The statistical significance is calculated from the difference in $\chi^2$
between the best fit and 
the fit with the signal yield fixed to zero.}
\label{tab_incl_fit}
\renewcommand{\arraystretch}{1.2}
\begin{ruledtabular}
\begin{tabular}{c|rcr}
& Yield, $10^3$ & Mass, $\mevm$ & Significance \\
\hline
$\Uo$     & $105.2\pm5.8\pm3.0$ & $9459.4\pm0.5\pm1.0$  & $18.2\,\sigma$\\
$\hb$     &  $50.4\pm7.8^{+4.5}_{-9.1}$ & $9898.3\pm1.1^{+1.0}_{-1.1}$  &  $6.2\,\sigma$\\
$3S\to1S$ &    $56\pm19$  & $9973.01$         &  $2.9\,\sigma$\\
$\Ut$     & $143.5\pm8.7\pm6.8$ & $10022.3\pm0.4\pm1.0$ & $16.6\,\sigma$\\
$\U(1D)$  &  $22.0\pm7.8$ & $10166.2\pm2.6$   &  $2.4\,\sigma$\\
$\hbp$    &  $84.4\pm6.8^{+23.}_{-10.}$ & $10259.8\pm0.6^{+1.4}_{-1.0}$ & $12.4\,\sigma$\\
$2S\to1S$ & $151.7\pm9.7^{+9.0}_{-20.}$ & $10304.6\pm0.6\pm1.0$ & $15.7\,\sigma$\\
$\Uth$    &  $45.6\pm5.2\pm5.1$ & $10356.7\pm0.9\pm1.1$ &  $8.5\,\sigma$
\end{tabular}
\end{ruledtabular}
\end{table}
The $\mmpp$ spectrum, after subtraction of both the combinatoric and $\ks\to\pp$ contributions
is shown with the fitted signal functions overlaid in Fig.~\ref{mmpp_all}.  
The signal parameters are listed in Table~\ref{tab_incl_fit}.

We studied several sources of systematic uncertainty.  
The background polynomial order was increased by three, and the range of the fits performed 
were altered by up to 100~$\mevm$.  
Different signal functions were used, 
including symmetric Gaussians and rCB functions with the width 
parameters left free.  We altered our selection criteria: tightening the requirements 
on the proximity of track origin to the IP, increasing the minimum number of tracks to four, 
and imposing 
the $\cos\theta_{\pp}<0.95$ 
requirement 
used in the $\uu\pp$ study.  In Table~\ref{tab_syst} a summary of our systematic studies is presented. 

\begin{table}[tb!h]
\caption{Absolute systematic uncertainties in the yields and masses
  from various sources. 
  }
\label{tab_syst}
\renewcommand{\arraystretch}{1.2}
\begin{ruledtabular}
\begin{tabular}{l|ccrc}
& Polynomial & Fit   & Signal & Selection    \\
& order      & range & shape  & requirements \\
\hline
$N[\Uo]$, $10^3$    & $\pm1.4$ & $\pm1.7$ & $\pm2.0$ &   --     \\
$N[\hb]$, $10^3$    & $\pm2.4$ & $\pm3.6$ & $^{+1.2}_{-8.0}$ &   -- \\
$N[\Ut]$, $10^3$    & $\pm3.4$ & $\pm3.2$ & $\pm5.0$ &   --     \\
$N[\hbp]$, $10^3$   & $\pm2.2$ & $\pm2.6$ & $^{+23.}_{-9.0}$ &   -- \\
$N[2\to1]$, $10^3$  & $\pm3.0$ & $\pm8.0$ & $^{+0}_{-18}$ &   --      \\
$N[\Uth]$, $10^3$   & $\pm1.0$ & $\pm3.0$ & $\pm4.0$ &   --     \\
\hline
$M[\Uo]$, $\mevm$   & $\pm0.04$ & $\pm0.06$ & $\pm0.03$ & $\pm0.18$ \\
$M[\hb]$, $\mevm$   & $\pm0.04$ & $\pm0.10$ & $^{+0.04}_{-0.20}$ & $^{+0.20}_{-0.30}$ \\
$M[\Ut]$, $\mevm$   & $\pm0.02$ & $\pm0.08$ & $\pm0.06$ & $\pm0.03$ \\
$M[\hbp]$, $\mevm$  & $\pm0.10$ & $\pm0.20$ & $^{+1.0}_{-0.0}$ & $\pm0.08$ \\
$M[2\to1]$, $\mevm$ & $\pm0.20$ & $\pm0.10$ & $\pm0.06$ & $\pm0.10$ \\
$M[\Uth]$, $\mevm$  & $\pm0.15$ & $\pm0.24$ & $\pm0.10$ & $\pm0.20$
\end{tabular}
\end{ruledtabular}
\end{table}
The values in the table represent the maximal change of parameters under the variations explored.
We estimate an additional $1\,\mevm$ uncertainty in mass measurements based on the 
differences between the observed values of the fitted $\U(nS)$
peak positions and their world averages. 
The total systematic uncertainties 
presented in Table~\ref{tab_incl_fit}
represent the sum in quadrature of all the contributions listed in Table~\ref{tab_syst}.
The signal for the $\U(1D)$ is marginal and therefore systematic uncertainties 
on its related measurements are not listed in the table. 
The significances of the $\hb$ and $\hbp$ signals, with systematic uncertainties accounted for, 
are $5.5\sigma$ and $11.2\sigma$, respectively.  

The measured masses of $\hb$ and $\hbp$ are $M=(9898.3\pm1.1^{+1.0}_{-1.1})\,\mevm$ and 
$M=(10259.8\pm0.6^{+1.4}_{-1.0})\,\mevm$, respectively.   Using
the world average masses of the $\chibnp$ states, we determine the hyperfine splittings to be
$\dmhf=(+1.6\pm1.5)\,\mevm$ and $(+0.5^{+1.6}_{-1.2})\,\mevm$,
respectively, where statistical and systematic uncertainties are combined in quadrature.  

We also measure the ratio of cross sections for $\ee\to\Uf\to\hbn\pp$ to that for $\ee\to\Uf\to\Ut\pp$.  
To determine the reconstruction efficiency we use the results of resonant structure studies reported in 
Ref.~\cite{belle_zb} that 
revealed the existence of two charged bottmonium-like states, $Z_b(10610)$ and $Z_b(10650)$, 
through which the $\pipi$ transitions we are studying primarily proceed.  
These studies indicate that the $Z_b$ most likely have
$J^P=1^+$, and therefore in our simulations the $\pipi$ transitions are generated 
accordingly.  To estimate the systematic uncertainty in our reconstruction efficiencies, 
we use MC samples generated with all allowed quantum numbers with $J\leq2$. 

We find that the reconstruction efficiency for the $\Ut$ 
is about 57\%, and that those for the $\hb$ and $\hbp$ relative to that
for the $\Ut$ are 
$0.913^{+0.136}_{-0.010}$ and 
$0.824^{+0.130}_{-0.013}$, respectively. 
The efficiency of the $R_2<0.3$ requirement is estimated from data by 
measuring signal yields with $R_2>0.3$. For $\Upsilon(2S)$, $\hb$ 
and $\hbp$ we find  $0.863\pm0.032$, $0.723\pm0.068$ and $0.796\pm0.043$, respectively.
From the yields and efficiencies described above, we determine the ratio of cross sections 
$R\equiv\frac{\sigma(\hbn\pp)}{\sigma(\Ut\pp)}$ to be
$R=0.46\pm0.08^{+0.07}_{-0.12}$ for the $\hb$ and
$R=0.77\pm0.08^{+0.22}_{-0.17}$ for the $\hbp$.  
Hence
$\Uf\to\hbn\pp$ and $\Uf\to\Ut\pp$ proceed at similar rates, despite the 
fact that the production of $\hbn$ requires a spin-flip of a $b$-quark. 

The rate of $\Uf\to\hbn\pipi$
is much larger than the upper limit for that of $\Uth\to\hbn\pp$ 
obtained by the \babar Collaboration~\cite{hb_pipi_babar}. 
This is consistent with the 
observation that the rates for $\Uf\to\U(mS)\pp$ with 
$m=1,2,3$ 
are much larger than those for $\U(nS)\to\U(mS)\pp$ for $n=2,3,4$~\cite{5s_rate}.  
The only previous evidence for the $\hb$ is a $3.0\sigma$ excess in $\Uth\to\piz\hb$ 
at $(9902\pm 4)\,\mevm$
presented by \babar~\cite{hb_babar}. 

We have also used $711\,\fb$ of $\ee$ collisions 
at the $\U(4S)$ resonance to search for $\U(4S)\to\hb\pp$ ($\hbp$ is kinematically forbidden).
The overall efficiency, assuming the $R_2$ efficiency at $\U(4S)$ to be the same as that at $\Uf$, 
is 
$0.94^{+0.11}_{-0.03}$ relative to that for $\Uf\to\hb\pp$.  From our observed yield of 
$(35\pm21^{+24}_{-15})\times10^3$, we therefore 
set an upper limit on the ratio of $\sigma(\ee\to\hb\pp)$
at the $\U(4S)$ to that at the $\Uf$ of
$0.27$ at 90\% C.L.

In summary, we have observed 
the $P$-wave spin-singlet bottomonium states $\hb$ and $\hbp$ in the reaction $\ee\to\Uf\to\hbn\pp$. 
The $\hbn$ masses correspond to hyperfine splittings that are consistent with zero.
We also have observed that the cross sections for these processes and that for
 $\ee\to\Uf\to\Ut\pp$ are of comparable magnitude, indicating 
the production of $\hbn$ at the $\Uf$ resonance must occur via 
a process that avoids the expected suppression related to heavy quark spin-flip.

We thank the KEKB group for excellent operation of the accelerator,
the KEK cryogenics group for efficient solenoid operations, and the
KEK computer group and the NII for valuable computing and SINET4
network support.  We acknowledge support from MEXT, JSPS and Nagoya's
TLPRC (Japan); ARC and DIISR (Australia); NSFC (China); MSMT
(Czechia); DST (India); MEST, NRF, NSDC of KISTI, and WCU (Korea);
MNiSW (Poland); MES and RFAAE (Russia); ARRS (Slovenia); SNSF
(Switzerland); NSC and MOE (Taiwan); and DOE and NSF (USA).

\end{document}